\documentclass[]{aastex701}

\begin{document}

\title{The Roman Space Telescope as a Planetary Defense Asset}

\author[0000-0002-6117-0164]{Bryan J. Holler}
\affiliation{Space Telescope Science Institute, Baltimore, MD, USA}
\email{bholler@stsci.edu}

\author[0000-0003-3047-615X]{Richard G. Cosentino}
\affiliation{Space Telescope Science Institute, Baltimore, MD, USA}
\email{rcosenti@stsci.edu}

\author[0000-0003-1796-9849]{William C. Schultz}
\affiliation{Space Telescope Science Institute, Baltimore, MD, USA}
\email{wschultz@stsci.edu}

\author[0000-0003-2630-8073]{Timothy D. Brandt}
\affiliation{Space Telescope Science Institute, Baltimore, MD, USA}
\email{tbrandt@stsci.edu}

\author[0000-0003-2638-720X]{Joseph R. Masiero}
\affiliation{Caltech/IPAC, Pasadena, CA, USA}
\email{jmasiero@ipac.caltech.edu}

\author[0000-0003-1383-1578]{Benjamin N. L. Sharkey}
\affiliation{Department of Astronomy, University of Maryland, College Park, MD, USA}
\email{sharkey@umd.edu}

\author[0000-0003-0743-9422]{Pedro H. Bernardinelli}
\affiliation{DiRAC Institute, Department of Astronomy, University of Washington, Seattle, WA, USA}
\email{phbern@uw.edu}

\author[0000-0002-4043-6445]{Carrie E. Holt}
\affiliation{Las Cumbres Observatory, Goleta, CA, USA}
\email{cholt@lco.global}

\begin{abstract}

NASA's Nancy Grace Roman Space Telescope, slated to launch in October 2026, will serve a critical role in the characterization and threat assessment of near-Earth Objects (NEOs), thus contributing to national and international planetary defense objectives. Operating from the Earth-Sun L2 point and observing in the near-infrared, Roman has the high sensitivity and high spatial resolution needed to measure the physical properties, compositions, and orbital trajectories of NEOs in order to understand their physical nature and potential hazards to Earth. Roman’s planetary defense capabilities complement those of two wide-field survey missions: the now operational ground-based Vera C. Rubin Observatory's Legacy Survey of Space and Time and the upcoming space-based NEO Surveyor. Rubin, observing in visible light, will discover over 100,000 NEOs. NEO Surveyor, observing in the mid-infrared where NEO thermal emission peaks, will detect 200,000--300,000 NEOs, some as small as $\sim$20 meters in diameter. With investment in developing the pipeline infrastructure required to extract information from moving target streaks, Roman will be able to observe NEOs down to the smallest sizes in order to improve our measurements of NEO orbits by 2--3 orders of magnitude, enable accurate diameter and albedo estimates in conjunction with NEO Surveyor, and reveal the spectral types and bulk compositions of the smallest NEOs. Together, these three US-led facilities will operate across the electromagnetic spectrum to form a comprehensive planetary defense network. 

\end{abstract}


\section{Introduction} \label{sec:Intro}
\setcounter{footnote}{0}

Launched to study the early universe and exoplanets closer to home, the Nancy Grace Roman Space Telescope (hereafter Roman) is NASA’s next flagship mission. Roman is designed with science objectives that push the boundaries of our understanding: mapping the elusive forces of dark energy and dark matter, charting the expansion history of the universe through precise measurements of supernovae, and building a statistical census of exoplanets to understand the demographics of planetary systems across our own galaxy.  

Achieving Roman's science objectives requires looking out through our own solar system, which opens up a remarkable opportunity. To map the distant universe, Roman must image vast swaths of the sky with high sensitivity and resolution, scanning through the space populated by near-Earth objects (NEOs) and other small solar system bodies. As a result, Roman is well-positioned to contribute not just to astrophysics, but to solar system and planetary science as well.

Roman shares a deeply complementary role in regards to solar system science to the Vera C. Rubin Observatory \citep[hereafter Rubin;][]{Ivezic2019}. Rubin, now fully online, recently released its first-look observations and identified $\sim$2000 new asteroids, several of which are NEOs. Like Rubin, Roman will detect NEOs as a natural result of its wide-field cosmology surveys, essentially capturing solar system small bodies ``for free'' while pursuing its core science objectives. 

But Roman is capable of more. When coordinated with Rubin and the upcoming NEO Surveyor mission \citep{Mainzer2023}, Roman can take a more active role in characterizing NEOs, leveraging its combined precise imaging and infrared capabilities. This synergy between missions, which is the focus of this paper, stands to enhance not only our understanding of the solar system’s NEO population, but also improve the broader planetary defense landscape. In Section~\ref{sec:NEOs} we provide an overview of NEOs and the history of surveys to detect and characterize them. Section~\ref{sec:RomanRubinNEOS} provides high-level overviews of these three missions while Section~\ref{sec:main} presents details on how Roman, in conjunction with Rubin and NEO Surveyor, can contribute to planetary defense in four key areas.

\section{Near Earth Objects} \label{sec:NEOs}

NEOs are asteroids or comets whose average orbital distance from the Sun is within 1.3 astronomical units (au). Within this population, asteroids are significantly more common than comets\footnote{\url{https://cneos.jpl.nasa.gov/}} and are broken into four sub-categories based on orbital properties and named after a prominent member of each group: Atiras, Atens, Apollos, and Amors. Planetary defense focuses on the Atens and Apollos, which have orbits that cross that of the Earth, thus presenting the potential for an impact. A further distinction can be made for potentially hazardous asteroids (PHAs), which are Atens or Apollos greater than 140 meters in diameter that have a minimum orbital intersection distance (MOID) with the Earth of $<$0.05 au.

NEOs are thought to originate as collisional fragments in the main asteroid belt, which then migrate towards unstable mean motion resonances with Jupiter via the Yarkovsky effect \citep[e.g.,][]{Morbidelli2002,Morbidelli2003} or the $\nu_6$ secular resonance with Saturn at $\sim$2 au \citep[e.g.,][]{Froeschle1986,Scholl1991}. Further interactions may alter their orbits, sending these asteroids towards the inner solar system. Their subsequent orbital lifetimes are only a few million years \citep[e.g.,][]{Morbidelli2002}, ending in impact or ejection from the solar system due to gravitational interactions with the Earth, Sun, or other inner solar system bodies.

Following Comet Shoemaker-Levy 9's impact into Jupiter in 1994 \citep[e.g.,][]{Hammel1995}, public awareness of NEO threats drastically increased. NEO threats even formed the plots of big-budget Hollywood movies including 1998's {\it Armageddon} and {\it Deep Impact}. Also in 1998, the United States Congress mandated that NASA identify and catalog 90\% of all NEOs larger than 1 km in diameter within 10 years\footnote{The denominator for the 90\% ``completeness'' limit is based on the number of objects estimated to be 1 km in diameter or larger using observed size-frequency distributions (SFDs; \citealt[e.g.,][]{Grav2023})}. This motivated the first targeted, large-sky surveys specifically designed and optimized to discover NEOs that could pose existential threats to humanity. In 2005, the George E. Brown Jr.~Near-Earth Object Survey Act\footnote{\url{https://www.congress.gov/committee-report/109th-congress/house-report/158/1}} directed NASA to detect, track, and catalog 90\% of NEOs down to 140 meters in diameter by 2020. This expanded the scope of the search dramatically by shifting the focus to smaller NEOs that would produce impacts with regional, not just global, consequences. However, this goal was not met by the 2020 deadline, nor has it been achieved at the time of writing, but it has served to motivate numerous ground- and space-based surveys for small NEOs.

Numerous ground-based NEO detection surveys followed, including the Lincoln Near Earth Asteroid Research (LINEAR) survey, Spacewatch, the Catalina Sky Survey, and the now-retired phase of Pan-STARRS \citep[e.g.,][]{Larson1998,Stokes2000,McMillan2007,Kaiser2002,Kaiser2010,Chambers2016}. Second-generation ground-based surveys include the Pan-STARRS NEO Survey, the Asteroid Terrestrial-impact Last Alert System (ATLAS), the Zwicky Transient Facility (ZTF), and the newly operational Vera C. Rubin Observatory, all of which continue to advance NEO detection and tracking \citep[e.g.,][]{Wainscoat2014,Tonry2018,Graham2019,Ivezic2019}. 

Ground-based NEO searches have focused on visible wavelengths. In space, the Wide-field Infrared Survey Explorer (WISE) and the NEOWISE mission extension \citep[e.g.,][]{Wright2010,Mainzer2011,Mainzer2014}, carried out observations at infrared wavelengths inaccessible from the ground. These infrared data combine with ground-based optical data to provide accurate size and albedo estimates for NEOs. WISE was especially critical for the discovery of darker objects invisible to optical surveys, but ended in 2024 after its orbit decayed and it entered Earth's atmosphere.

\section{Next Generation Survey Facilities} \label{sec:RomanRubinNEOS}

The now-operational NSF–DOE Vera C.~Rubin Observatory will play a major role in discovering and tracking NEOs as part of its Legacy Survey of Space and Time \citep[LSST;][]{Ivezic2019}. Located at Cerro Pach\'on in Chile, the facility features an 8.4-meter primary mirror and a wide-field optical camera with a 9.6 deg$^2$ field of view, covering wavelengths from 320 to 1050 nanometers across six optical bands ({\it ugrizy}). Rubin's field of regard is limited by Earth’s horizon and weather, but optimized to revisit most survey areas every 3–-4 nights. The main survey will image about 18,000 deg$^2$ of sky, primarily in the southern hemisphere and along the ecliptic. NEO detection is achieved through frequent, time-separated exposures and difference imaging, enabling the identification and orbit tracking of fast-moving, faint objects in the visible spectrum. Early public-release images from the Rubin Observatory, representing 1.4$\times$ the telescope’s field of view, identified over 2100 new small bodies, including seven new NEOs. Rubin is predicted to discover $\sim$127,000 new NEOs by the end of the 10-year LSST \citep{Kurlander2025}.

NEO Surveyor, scheduled for launch in 2027, is NASA’s dedicated infrared NEO discovery mission developed in direct response to the unachieved 2005 Congressional mandate. Its primary objective is to detect $>$90\% of NEOs larger than 140 meters in diameter, particularly dark, low-albedo objects, by measuring their thermal infrared emission {bf \citep{Mainzer2023}}. Operating from the Sun–Earth L1 Lagrange point, NEO Surveyor will have a stable, space-based vantage point with access to regions of the sky near the Sun that cannot be observed from Earth. The spacecraft carries a 50-cm telescope equipped with infrared detectors sensitive to two thermal infrared bands: 4–5.2 $\mu$m and 6–10 $\mu$m. Its field of regard spans solar elongation (Sun–observer–target) angles from approximately 45 to 120$^{\circ}$ and ecliptic latitudes between $\pm$40$^{\circ}$. NEO Surveyor will continuously scan this region with a $\sim$13-day cadence, obtaining multiple detections of each object to enable both orbit determination and thermal characterization.

Currently, nearly 95\% of NEOs larger than 1 km in diameter have been identified, achieving the 1998 Congressional goal. However, only about 40\% of 140-m-class NEOs are believed to be known, based on estimated size–frequency distributions \citep{Grav2023}. NEO Surveyor is specifically designed to close this gap for larger NEOs while also providing statistical information on smaller, largely unknown populations of 10s-of-meters-sized objects, such as the $\sim$20-m asteroid that caused localized destruction in Chelyabinsk, Russia, in February 2013.

The Roman Space Telescope brings unique and complementary capabilities to the Rubin and NEO Surveyor missions through its combination of high spatial resolution, stable space-based imaging, and wide-field infrared sensitivity \citep[e.g.,][]{Holler2018,Akeson2019,Schlieder2024}. Roman's Wide Field Imager (WFI) provides a wide field of view (0.281 deg$^2$), exquisite detector sensitivity, and an angular resolution $\approx$10 times better than Rubin. These capabilities will enable precise astrometric follow-up of faint, moving objects discovered by Rubin, and provide critical near-infrared observations that bridge the gap between Rubin’s optical detections and NEO Surveyor's thermal infrared measurements. During Roman's 5-year primary mission, it will dedicate 74.5\% of the total time to predefined Core Community Surveys (CCS) and 25.5\% to General Astrophysics Surveys (GAS) submitted by the community in response to regular calls for proposals \citep{ROTAC2025}. Each Roman CCS focuses on different mission-level science goals. The High-Latitude Wide-Area Survey (HLWAS) will cover over 5000 square degrees of the sky to different depths, with no return visits, while the High-Latitude Time-Domain Survey (HLTDS) and Galactic Bulge Time-Domain Survey (GBTDS) rely on repeat observations of smaller portions of the sky at specific, defined cadences. The HLWAS includes coverage over a wide range of ecliptic latitudes, including the ecliptic itself, and is therefore the survey that will be most useful for serendipitous observations of NEOs and other small bodies. Targeted observations, likely as part of a GAS, that revisit survey fields from Rubin and NEO Surveyor with consistent sensitivity and without atmospheric interference make Roman ideal for refining orbits, characterizing surface properties, and identifying small body populations in crowded or high-background regions of the sky. 

Together, Rubin, NEO Surveyor, and Roman form a complementary planetary defense network. Their combination of wide-field optical searches, space-based infrared detection, and high-precision follow-up provides unmatched capability in finding, tracking, characterizing, and assessing the threat posed by NEOs.

\section{Roman's Planetary Defense Capabilities} \label{sec:main}

In this section, we present Roman's ability to characterize the threat posed by NEOs. Studies of small solar system bodies follow a common pattern, naturally beginning with discovery and then orbit determination. Once the orbit is known well enough to confidently conduct follow-up observations, the next steps focus on physical and spectral characterization. The same is true for NEOs from a planetary defense perspective, just on a more compressed timeline and with much higher stakes. NEO Surveyor and Rubin will operate at cadences that will enable discovery of NEOs, which requires multiple observations over a short time period to compute an initial orbit. There are no plans for Roman to operate at this same cadence, so its contribution to planetary defense arises from its ability to follow up and characterize newly discovered NEOs. In this section we first discuss the range of NEOs that Roman will be capable of observing, then focus on three areas in which Roman can make significant contributions to planetary defense: orbit improvement, diameter determination, and spectral characterization. We describe how Roman complements Rubin and NEO Surveyor in these three areas.

It is important to keep in mind that the capabilities of Roman described in the remainder of this paper cannot currently be realized due to the way that the calibration pipeline treats moving targets. Roman does not have moving target tracking capabilities: all tracking will be sidereal using guide stars imaged on the WFI detectors. As a result, all solar system objects will produce streaks across one or more detectors. The default Roman calibration pipeline will flag the majority of these streaks as cosmic rays and remove them from the images. Solar System science and planetary defense objectives will therefore be unachievable with Roman unless effort and resources are dedicated to develop the means to recover and analyze moving target signatures in WFI images.

\subsection{Observing Smaller and Faster NEOs}

Identification is the first, and arguably most important, step in evaluating the threat a NEO may pose to the Earth. NEO Surveyor was developed to discover NEOs as small as 20--25 meters in diameter by targeting their peak thermal emission. These mid-infrared wavelengths are blocked by Earth's atmosphere and can only be observed from space. Roman and Rubin operate at shorter wavelengths, targeting the reflected sunlight of a NEO rather than its thermal emission. Roman and Rubin have overlapping wavelength coverage from $\sim$0.5--1.0 $\mu$m but Roman is capable of observing smaller and fainter objects than Rubin. The Rubin 5-$\sigma$ limiting magnitude\footnote{\url{https://smtn-002.lsst.io/}} in a single 30-second exposure is $r$=24.52. Using the Roman Exposure Time Calculator\footnote{\url{https://roman.etc.stsci.edu/}}, we calculated that the 5-$\sigma$ detection threshold is reached in 60-, 300-, and 980-second exposures for limiting magnitudes of $r$=25.18, 26.47, and 27.13, respectively, for the wide-band F146 filter (0.927--2.0 $\mu$m).

\begin{figure}[h!]
\centering
\includegraphics[width=0.6\columnwidth]{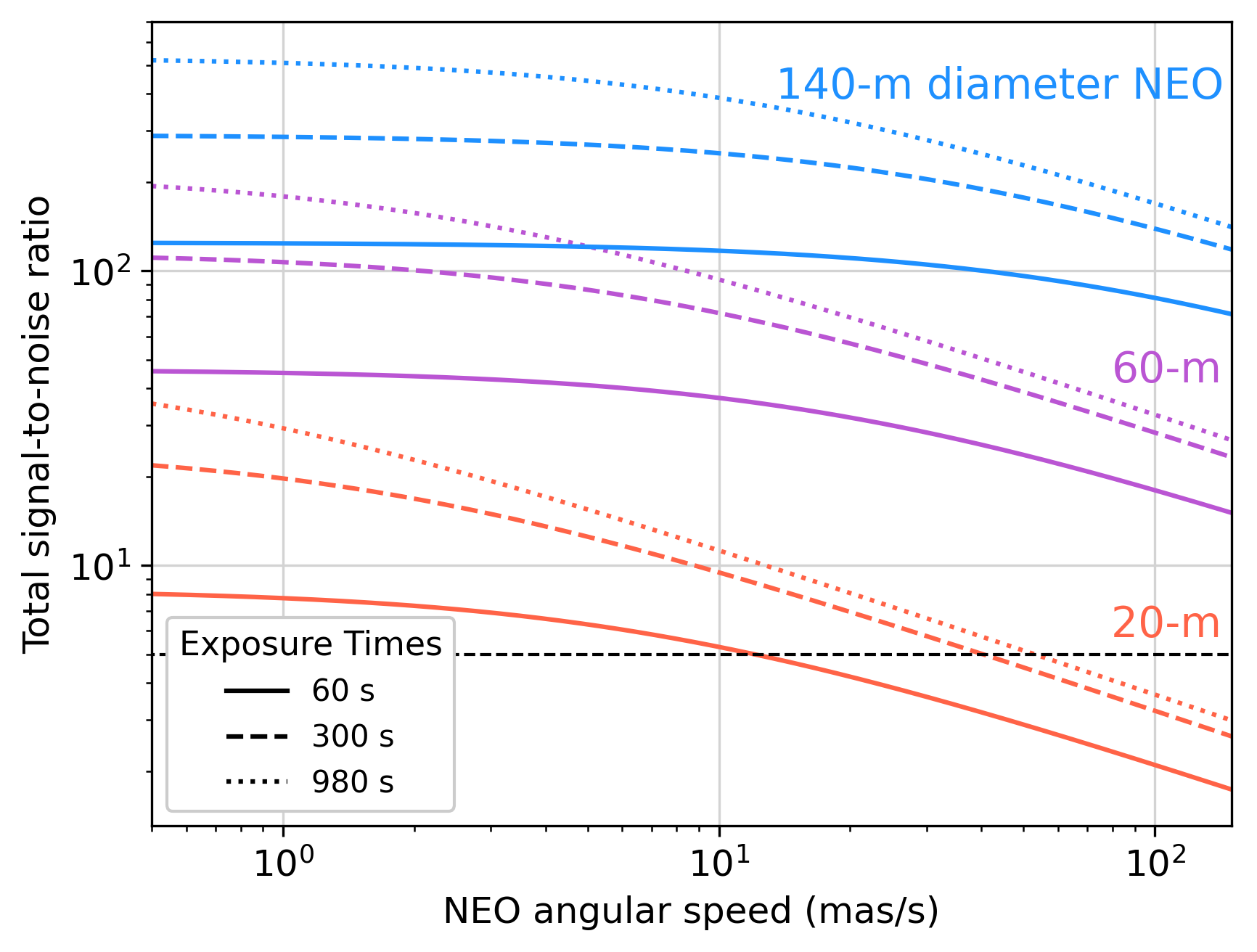}
\caption{Roman has the capability to detect NEOs with a total signal-to-noise ratio of 5 (dashed black line) for all objects in the $\sim$20--140 m size range for a wide range of angular speeds. Detection of NEOs below the expected NEO Surveyor threshold (i.e., $<$20 meters) is also possible, but only at lower angular speeds. \label{fig:detection_angrate}}
\end{figure}

To investigate how the motion of the targets affects their identification in images, we generated thermal models for NEOs with diameters of 20, 60, and 140 meters, assuming visible geometric albedos of 8\%, a heliocentric distance of 1.1 au, a solar elongation angle of 92$^{\circ}$ (the middle of Roman's field of regard), an emissivity of 0.9, and a beaming parameter of 0.756. (See Section~\ref{sec:sizedeterm} for additional information on the computation of thermal models.) The diameters reflect the size range of NEOs targeted by NEO Surveyor (20--140 meters; \citealt{Mainzer2023}). For additional context, the former ``city killer'' asteroid 2024 YR$_4$ is 60 meters in diameter \citep{Rivkin2025}, and this may also have been the size of the putative Tunguska object. The models were uploaded to the Roman ETC and used to determine the total number of electrons obtained in a 9-pixel extraction aperture centered on a fixed point source observed with the F146 filter. We then computed the linear motion of the NEOs for three different fixed exposure time options (60, 300, and 980 seconds) over a range of apparent angular speeds (0.5--150 milliarcseconds per second, or mas/s). Based on daily angular speeds retrieved from JPL Horizons over a 5-year period for $\sim$850 NEOs, angular speeds largely fall in the $\sim$1--100 mas/s range when a NEO is in the Roman field of regard (solar elongation angles between 54$^{\circ}$ and 126$^{\circ}$). For our exploration, we assumed a constant angular speed and uniform distribution of counts across pixels, such that the area of the streak increased linearly. The combined signal-to-noise ratio of the resulting streak was computed by assuming a noise model that included detector read noise, Poisson noise from the target itself, and Poisson noise due to background removal added in quadrature. For simplicity, we assumed a read noise of 10 electrons for each exposure length\footnote{See Instrumental Noise article in the Roman Documentation (RDox).}. 

In the limit of low angular speed, the detectability of a NEO is the same as that of a star of the same brightness. At high angular speeds, the NEO's signal is spread out over $n$ pixels, where $n$ is the product of the integration time $t$ and the angular speed $v$ in units of pixels/s (a WFI pixel is $\approx$110 mas). Assuming that we are background-limited, the signal-to-noise ratio in each pixel decreases with $v$. Summing the signal over the entire track increases the signal-to-noise ratio with $\sqrt{n} \propto \sqrt{v}$. The net result is that, at high angular speeds, the signal-to-noise ratio scales as $1/\sqrt{v}$. Realizing this sensitivity in practice will require fitting the track in the individual Roman/WFI reads as the detector is read out up-the-ramp.

Fig.~\ref{fig:detection_angrate} demonstrates that Roman is capable of observing any NEO down to a few 10s of meters over the full range of angular speeds examined. Such targets can be observed by Roman whenever they are within the field of regard. NEOs in the 20-meter size range are more easily detected with longer exposure times, but even slower-moving objects ($\sim$10 mas/s) can be observed $\sim$30\% of the time in the shortest Roman exposure of 60 seconds. This figure also indicates that NEOs smaller than 20 meters can be seen by Roman, albeit at lower angular speeds, which exceeds the smallest diameter expected to be observed by NEO Surveyor.

\subsection{Improving NEO Orbits}

The orbital trajectory of a newly-discovered NEO determines whether it is a PHA that poses a near-term threat to Earth. NEO Surveyor will excel at discovering NEOs, both interior and exterior to the Earth's orbit, and will provide initial orbit estimates, while Roman will have the higher precision needed to refine those orbits and identify threats. NEO Surveyor is expected to discover $\sim$200,000--300,000 new NEOs during its 5-year baseline mission, increasing the number currently known by an order of magnitude \citep{Mainzer2023}. While the vast majority will be harmless, a few will merit rapid and detailed follow-up observations to perform a risk assessment.

\begin{figure}[h!]
\centering
\includegraphics[width=0.9\columnwidth]{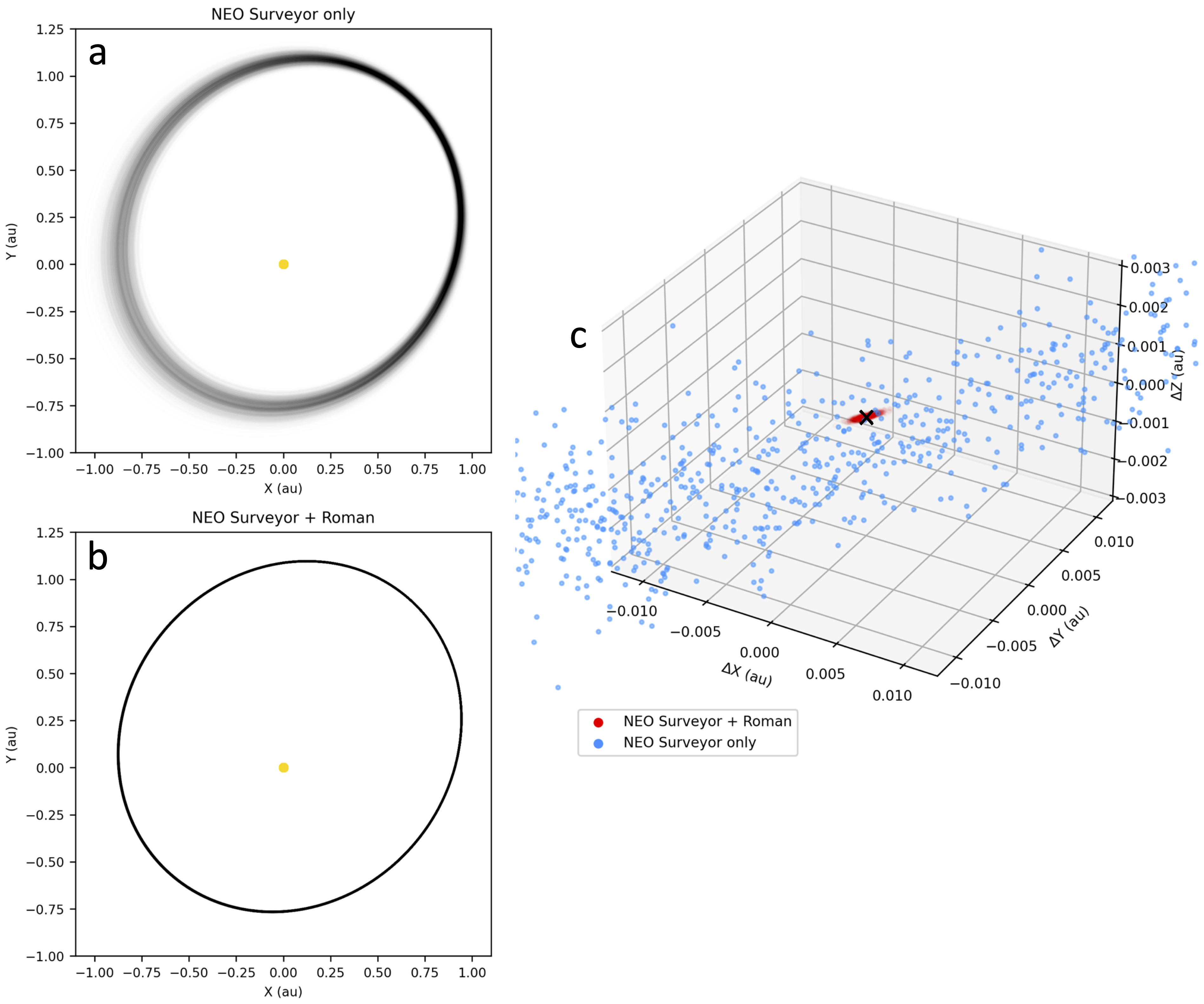}
\caption{Monte Carlo simulated orbit fits for the NEO Aten assuming NEO Surveyor astrometry spaced 13 days apart (a) and NEO Surveyor astrometry followed by Roman astrometry 7 days later (b). The yellow dot shows the position of the Sun (not to scale). (c) Orbit clone distributions one month after the last astrometric data point for the NEO Surveyor only (blue) and NEO Surveyor and Roman (red) scenarios. The black `x' is the mean position of both distributions. A single additional astrometric observation of an NEO by Roman decreases the orbital uncertainty by 95-99\% up to a year later.
\label{fig:orb_char}}
\end{figure}

We modeled a NEO orbit to quantify the power of Roman astrometry when following up on a NEO Surveyor discovery. The default NEO Surveyor survey pattern is to conduct repeat visits to the same patch of sky four times in a 6--9-hour period (we refer to this set of four visits as an ``epoch'' hereafter), with the entire field of regard requiring $\sim$13 days to cover \citep{Mainzer2023}. More than one observation of a NEO is required to create a model of its orbit, so we investigated two scenarios: one where only one NEO Surveyor baseline (consisting of two epochs) is obtained, and another where a Roman follow-up observation occurs one week after a NEO Surveyor baseline. We used JPL Horizons\footnote{\url{https://ssd.jpl.nasa.gov/horizons/app.html\#/}} to retrieve the sky coordinates (right ascension and declination) of the known Earth-crossing NEO, Aten, at the observing cadence of NEO Surveyor. We then conducted 1000 Monte Carlo trials for each scenario with Gaussian-like perturbations to the orbit, simulating the time between observations and assuming astrometric uncertainties of 10\% of a pixel (i.e., 0.3$''$ for NEO Surveyor and 0.011$''$ for Roman). Each trial included an orbit fit using the {\tt LambertFit}\footnote{\url{https://github.com/bengebre/lambertfit}} Python routine, which uses the provided astrometry to compute the Cartesian (X, Y, Z) positions of the object at the start and end of the time range. We calculated the semi-axes of the resulting error ellipsoid as the standard deviations in the X, Y, and Z coordinates at the end of each scenario. We then propagated the orbit forward in time by 7, 30, and 365 days after the last astrometric measurement and recalculated the semi-axes of the error ellipsoid in each case. A more detailed discussion of this process is presented in Appendix~\ref{sec:orbit_calc}. The orbit fits for both scenarios are presented in Fig.~\ref{fig:orb_char}. 

We found that including a single Roman data point one week after a NEO Surveyor baseline reduced the uncertainties in the X, Y, and Z coordinates by 95--99\% up to a year after the last astrometric measurement. Part of this improvement may arise from the triangulation achieved with observations from both the Earth-Sun L1 and L2 Lagrange points. Fig.~\ref{fig:orb_char}c presents the distributions of the 1000 simulated Aten clones one month after the last astrometric measurement for each scenario. The reduction in orbital uncertainties is such that, if Earth were placed at the mean position of the two distributions, (0, 0, 0) in the plot, the percentage of clones that pass within the cross-section of the Earth (accounting for gravitational focusing as described in, e.g., \citealt{Hughes2003}) increases from 0\% to 1.4\%. While this does not seem like a large change, it would be enough to increase the impact probability above the 1\% threshold for the International Asteroid Warning Network (IAWN) to issue an official warning. The scenario described here would give governments more time to react and underscores the importance of fast follow-up with Roman. When considering the value of an early warning for something like the Chelyabinsk event in February 2013, which was caused by an object much smaller than a ``city killer,'' people could have had the chance to secure their homes, businesses, and vehicles against damage and to shelter in place away from windows, thereby preventing hundreds of injuries.

\begin{figure}[h!]
\centering
\includegraphics[width=0.5\columnwidth]{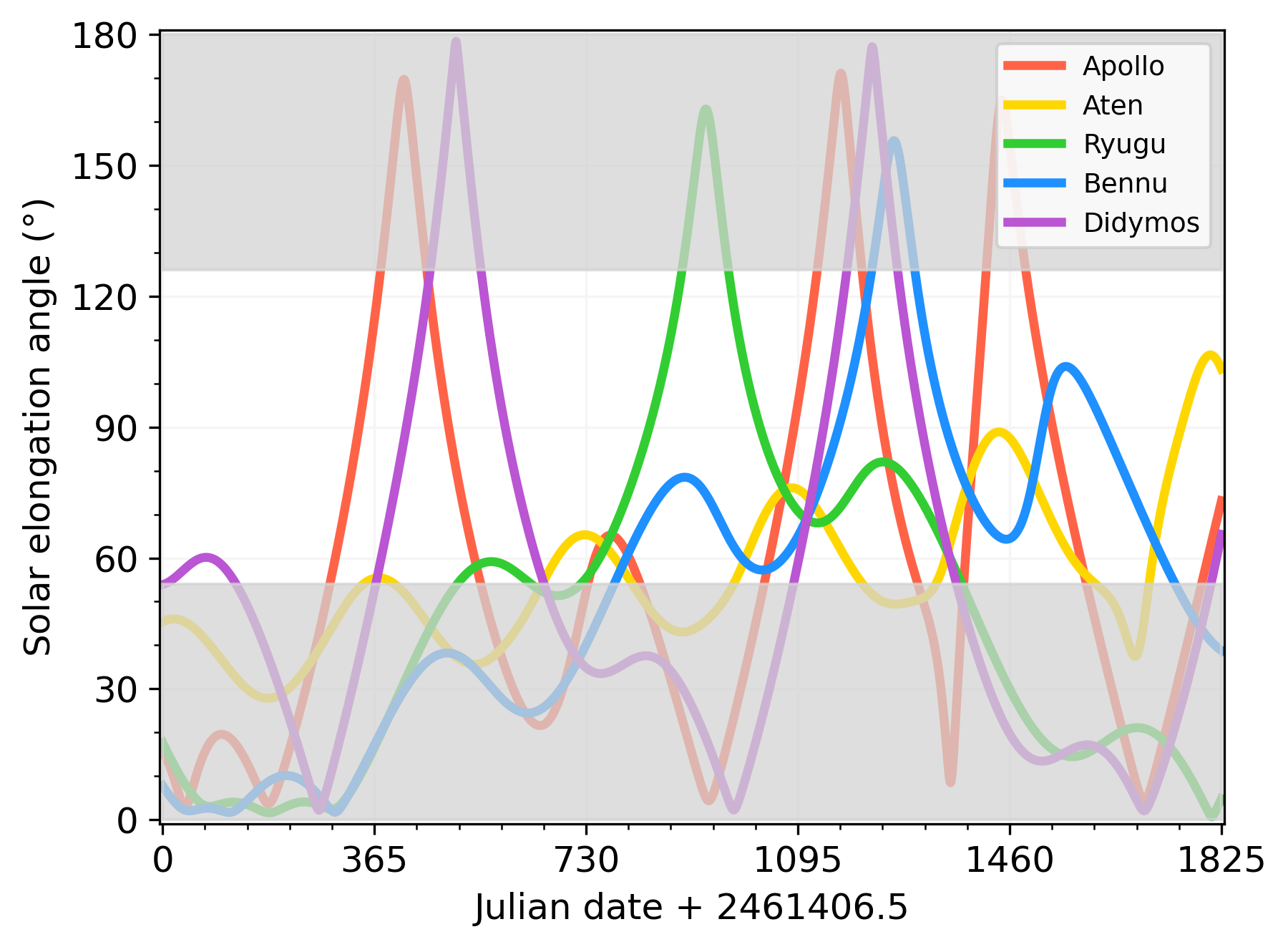}
\caption{Solar elongation (Sun-observer-target) angles for five representative NEOs over the course of 5 years starting on January 1, 2027. Roman's field of regard spans 54--126$^{\circ}$ and the grey shaded regions indicate unobservable elongation angles. Due to these viewing geometry constraints, some NEOs cannot be observed by Roman for multiple consecutive years, which increases the importance of rapid astrometric follow-up. \label{fig:solarelong}}
\end{figure}

Furthermore, observability of NEOs for space-based facilities with restricted fields of regard, such as Roman and NEO Surveyor, does not follow a regular pattern. Figure~\ref{fig:solarelong} presents the solar elongation (Sun-observer-target) angle for five representative NEOs over a 5-year period starting on January 1, 2027. As seen in the figure, some objects can experience extended periods of time without being observable. For instance, Bennu cannot be observed by Roman for over two years from late 2026 to early 2029. Ground-based facilities, such as Rubin, can only observe at elongation angles $\gtrsim$90$^{\circ}$, while NEO Surveyor covers a similar range to Roman (45--120$^{\circ}$). Thus, fast follow-up may not only be beneficial, but crucial for reducing orbital uncertainties for newly discovered NEOs before they become unobservable, especially for those below the detection limit of ground-based telescopes.

This section focused primarily on the synergies between Roman and NEO Surveyor, but Roman may also be able to make contributions for NEOs discovered by Rubin. Granted, the smaller pixel size and faster observational cadence of Rubin would likely make such contributions smaller. However, only 30--40\% of Rubin-discovered NEOs with diameters $<$140 meters are expected to have observational arcs longer than one year at any point during the LSST duration, with even shorter arcs for smaller NEOs \citep{Kurlander2025}; Roman's ability to contribute meaningful astrometric measurements therefore increases with decreasing diameter. NEO Surveyor will almost certainly observe these objects as well, thus increasing the number and time baseline of astrometric observations, with Roman capable of providing the final, higher-precision measurement in cases where the impact probability remains elevated.

\subsection{Determining Sizes with Higher Precision} \label{sec:sizedeterm}

As discussed previously, the definition of a ``potentially hazardous asteroid'' has two components: orbit and diameter. Determination of a NEO's diameter is more challenging than computing the orbit, requiring imaging observations across a wider wavelength range. We demonstrate this, and the contribution of Roman, using the Near-Earth Asteroid Thermal Model (NEATM; \citealt[e.g.,][]{Harris1998}). The NEATM takes as inputs the diameter, albedo, heliocentric distance, observer distance, emissivity (thermal emission efficiency), and beaming parameter (a measure of the surface roughness). The output is a spectrum representing the reflected light and thermal emission from the modeled object.

\begin{figure}[h!]
\centering
\includegraphics[width=0.45\columnwidth]{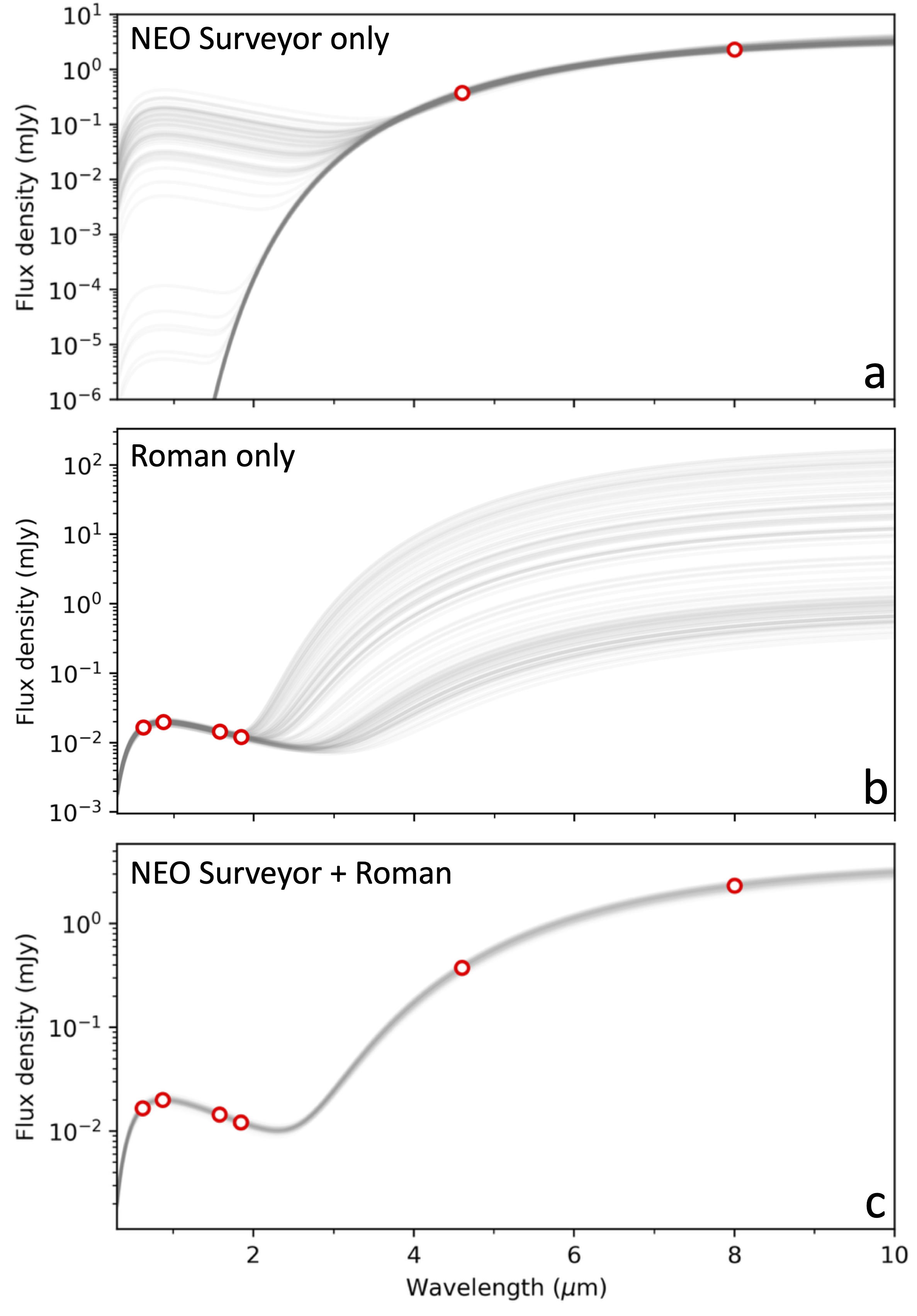}
\caption{Thermophysical model fits for (a) NEO Surveyor photometry only, (b) Roman photometry only, and (c) combined NEO Surveyor and Roman photometry. 1000 Monte Carlo trials (grey lines) were performed using an implementation of the Near-Earth Asteroid Thermal Model (NEATM; \citealt[e.g.,][]{Harris1998}). Red points are the photometric measurements (error bars are smaller than the symbols). Neither Roman nor NEO Surveyor by themselves are capable of obtaining the measurements needed to accurately determine the diameter and albedo simultaneously, only the combination of the two facilities is able to accomplish this goal. \label{fig:diameter}}
\end{figure}

We generated a model for a 140-meter asteroid with a visible geometric albedo of 8\%, a heliocentric distance of 1.1 au, an observer distance of 0.425 au (assuming a solar elongation angle of 92$^{\circ}$), and standard values of 0.9 and 0.756 for the emissivity and beaming parameters, respectively. We calculated the modeled flux density (in mJy) at the central wavelength of four Roman filters and the two NEO Surveyor filters. The Roman filters considered were F062 (0.620 $\mu$m), F087 (0.869 $\mu$m), F158 (1.577 $\mu$m), and F184 (1.842 $\mu$m), which provide the best spectral characterization, as described in Section~\ref{sec:spec_char}. NEO Surveyor operates over two wavelength ranges, 4.0--5.2 $\mu$m and 6.0--10.0 $\mu$m, so we used central wavelengths of 4.6 $\mu$m and 8 $\mu$m, respectively. Uncertainties (1-$\sigma$) on the flux densities were assumed to be 10\% (a signal-to-noise of 10). Then 1000 Monte Carlo trials were run where the photometry in each filter was allowed to vary within the 1-$\sigma$ error bars for three different scenarios: NEO Surveyor only, Roman only, and combined NEO Surveyor and Roman. In each trial, the best-fit NEATM model was computed and the diameter and visible geometric albedo were recorded. The beaming parameter, which is typically treated as a free parameter in the NEATM model, was fixed at 0.756, the same value used to generate the original model, in order to evaluate the retrieval of the diameter and albedo. The nominal diameter and albedo were computed as the mean of the 1000 trials and the uncertainty as the standard deviation.

Making use of only the NEO Surveyor photometry (Fig.~\ref{fig:diameter}a) resulted in an adequate recovery of the diameter (149$\pm$20 m) but a poor determination of the albedo (16$\pm$19\%); note that albedo cannot be negative. Removing the albedo values $<$3\% resulted in an increase to the albedo estimate to 29$\pm$16\%, which does not include the true value of 8\% within the 1-$\sigma$ uncertainty. However, the diameter estimates are accurate to within 1-$\sigma$. The Roman only scenario (Fig.~\ref{fig:diameter}b) produced less accurate and less precise values for both the diameter (274$\pm$256 m) and albedo (15$\pm$15\%). On the other hand, the combination of NEO Surveyor and Roman photometry (Fig.~\ref{fig:diameter}c) recovered the initial inputs for diameter (140$\pm$5 m) and visible geometric albedo (8.0$\pm$0.7\%) to high precision. Both facilities are needed to accurately and precisely constrain the diameters and albedos of NEOs because of the degeneracy between these two quantities. For instance, looking only in the near-infrared, a large, low-albedo object and a small, high-albedo object can produce similar fluxes; thermal measurements in the mid-infrared are needed to discern between the two scenarios. As such, observations in the visible or near-infrared are required to better constrain the visible geometric albedo (calculated at 0.55 $\mu$m) and observations in the mid-infrared are required to better constrain the diameter. Roman is more than capable of providing the necessary short-wavelength spectral leverage to complement NEO Surveyor observations at longer wavelengths, especially for the smaller NEOs that cannot be easily observed from the ground.

\subsection{Revealing Bulk Compositions}\label{sec:spec_char}

\begin{figure}[h!]
\centering
\includegraphics[width=0.8\columnwidth]{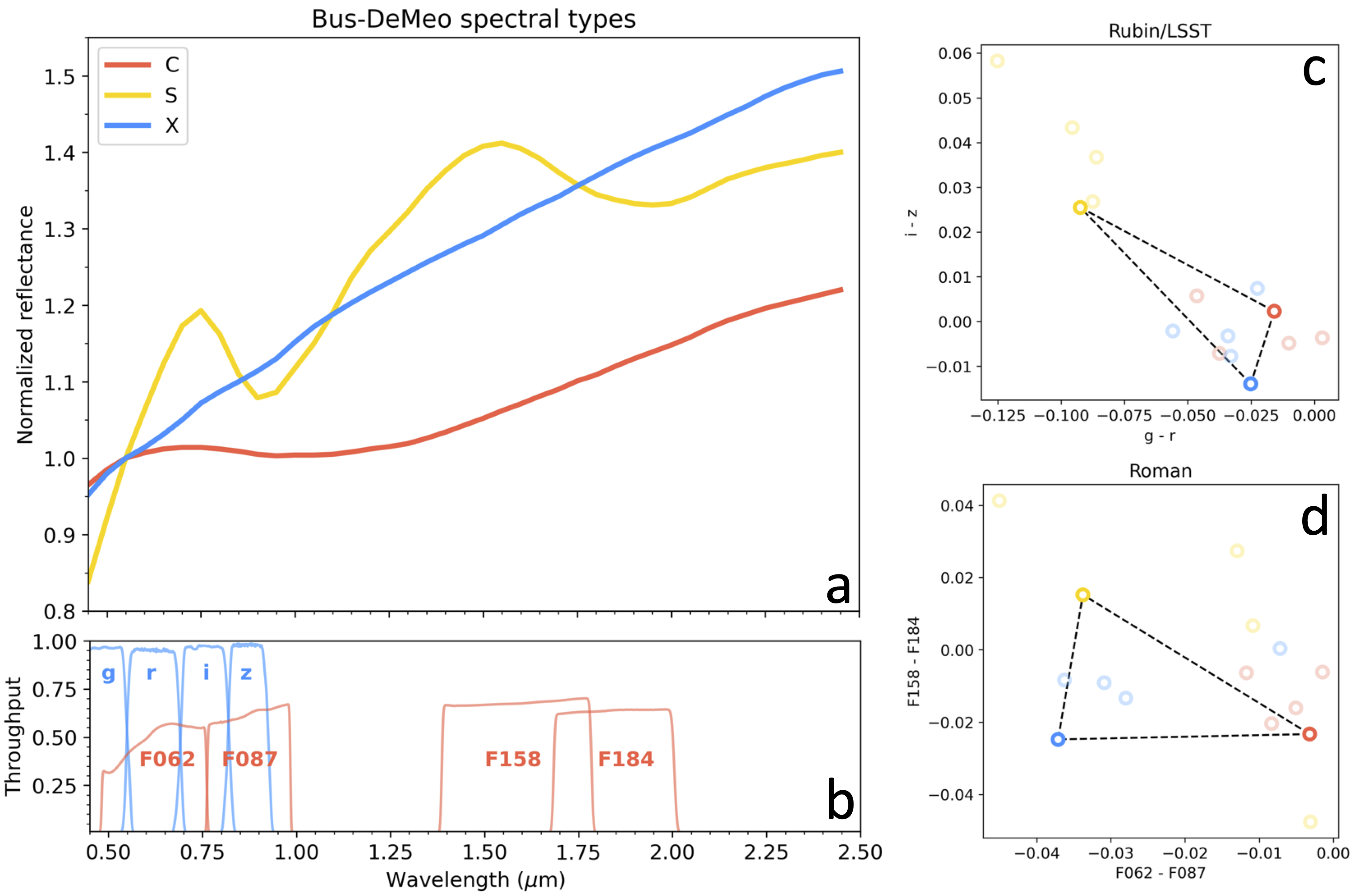}
\caption{(a) Bus-DeMeo spectral archetypes for the three most common asteroid spectral types: C, S, and X \citep{DeMeo2009}. (b) Filter throughput profiles for Rubin (blue) and Roman (red). (c) \& (d) Colors of the three most common asteroid spectral types (same colors as panel a) in the Rubin/LSST and Roman photometric systems, respectively. Lighter points indicate asteroid sub-types within the C, S, and X complexes. The largest average color-color Euclidean distance between points is presented for each mission. The two missions have comparable capabilities for sorting NEOs by spectral type, but only Roman can perform this task for the smallest observable NEOs ($\sim$20--30 meters). \label{fig:busdemeo}}
\end{figure}

Asteroid compositions are commonly characterized by their visible and near-infrared spectra using the Bus-DeMeo taxonomy \citet{DeMeo2009}. The dominant spectral types among main belt asteroids (Fig.~\ref{fig:busdemeo}) are the silicate-rich (S-complex), carbonaceous (C-complex), and a degenerate spectral class that includes a wide variety of compositions that includes metal-rich objects (X-complex). Most NEOs are thought to derive from the main belt, and their spectral distribution spans a similar broad range, with a notably larger fraction of S-complex objects \citep{Binzel2019,Sanchez2024}. Differentiating between specific taxonomic subgroups requires spectroscopic observations to detect mineral species, but rough differentiation between S-, C-, and X-types is possible using photometric colors. From a planetary defense perspective, determining the taxonomy of an impactor provides an important constraint on its material strength and density, which when combined with diameter and albedo constrains the object’s total mass and porosity. Some compositional information is provided in a statistical sense by albedo measurements alone, but there is substantial degeneracy between the albedos of S types \citep[$\sim$0.2;][]{DeMeo2013} and the range observed across the metal-rich asteroids \citep[0.12--0.23;][]{Shepard2010}. Differentiating between S, C, and X color classes therefore refines target characterization to inform key risk assessment parameters \citep[][see their Fig. 3]{Dotson2024} and assists the planning of intensive characterization campaigns for targets of interest \citep[e.g.,][]{Reddy2019,Reddy2022a,Reddy2022b,Reddy2024}

We investigated the ability of Rubin and Roman to sort NEOs into these spectral groups using the spectral archetypes from \citet{DeMeo2009} and the respective filter sets of each facility. The archetypes were reflectance spectra normalized to unity at 0.55 $\mu$m, so the first step was to multiply them by a blackbody at 5772 K (the temperature of the Sun) to obtain flux units. We then performed synthetic filter photometry by convolving the filter throughput profiles for Rubin (only the $griz$ filters, which account for the majority of detections per \citealt{Kurlander2025}) and Roman (all filters) with the asteroid spectra and integrating the area under the curve. The same was done for the blackbody to obtain an estimate of the solar flux in each bandpass.

Every possible color-color combination was examined within the Rubin and Roman filter sets, using the solar flux in each bandpass to adjust the values with respect to the Sun. The color-color combination that best separated the three spectral types was determined by calculating the average Euclidean distance between every set of points and identifying the maximum value (Fig.~\ref{fig:busdemeo}). Sub-classes within each complex are shown in Fig.~\ref{fig:busdemeo} for completeness but were not included in the Euclidean distance calculation. For Rubin, this was $i-z$ vs $g-r$ and for Roman this was F158$-$F184 vs F062$-$F087. Near-infrared spectra are ultimately required to confirm the spectral type, but Roman and Rubin will both be capable of providing an initial spectral characterization. This is important to quickly infer an approximate bulk (internal) composition without using time-intensive spectroscopic follow-up facilities, such as JWST \citep[e.g.,][]{Rivkin2025}. Constraining the composition and albedo, which in turn helps constrain the diameter, leads directly to an estimate of the magnitude of a potential impact \citep{Dotson2024}. Roman will be able to carry out this initial characterization for the smallest observed NEOs ($\sim$20--30 meters in diameter), which will largely be observable by Rubin only in the $r$ and $i$ bands.

\section{Summary}

Effective planetary defense that protects lives, property, and business interests requires coordination across a ground- and space-based network of state-of-the-art telescopic facilities. In addition to the large-sky surveys currently in operation, the next few years will see the initiation of the Vera C. Rubin Observatory's Legacy Survey of Space and Time and the launch of the NEO Surveyor and Nancy Grace Roman Space Telescope missions. These three next-generation survey facilities, all led by the United States, will provide unprecedented detection and characterization capabilities for evaluating the threats that newly discovered NEOs may pose to Earth. Starting in early 2027, Roman will be positioned to fit seamlessly into this planetary defense network and make immediate contributions to the characterization of potentially hazardous near-Earth objects, by specifically:

\begin{itemize}

\item Observing NEOs as small as tens of meters moving at any angular speed for follow-up characterization.

\item Decreasing orbital uncertainties by 95-99\% with a single astrometric measurement one week after initial discovery by NEO Surveyor, thereby helping identify high-priority targets in need of coordinated follow-up observations.

\item Adding near-infrared photometry to thermal infrared measurements from NEO Surveyor to improve the diameter and visible geometric albedo estimates and inform impact models.

\item Categorizing newly discovered NEOs by spectral type, thus revealing their internal compositions for evaluating impact scenarios.

\end{itemize}

Of the three facilities, only Roman was designed without planetary defense in mind. A NEO pilot survey with Roman during its first year of operations, prior to the launch of NEO Surveyor and well into the Rubin LSST, should utilize the strategic capabilities we have introduced. Our recommendation would be for the Roman mission to execute an optimized gap-filling survey, between observations of astrophysics surveys, that would specifically target the ecliptic plane utilizing the diagnostic filter set. Such a data set would detect numerous NEOs of all sizes, angular speeds, and spectral types and unlock the potential of the strategic synergies presented in this paper. A plan for early observations of NEOs, combined with modifications to the calibration pipeline necessary to recover moving target streaks, will enable Roman to contribute to the new era of planetary defense from day one.

\begin{acknowledgments}
We acknowledge financial support from the STScI Director’s Research Funds (DRF).
\end{acknowledgments}

\begin{contribution}
BJH led the quantitative analysis. BJH, RGC, and WCS contributed equally to the construction and editing of the manuscript. TDB, JRM, BNLS, PHB, and CEH contributed edits to the manuscript.
\end{contribution}

\appendix

\section{Conversion between state vectors and orbital elements} \label{sec:orbit_calc}

The conversion between state vectors, ($X$, $Y$, $Z$, $v_X$, $v_Y$, $v_Z$), and orbital elements, ($h$, $e$, $i$, $\Omega$, $\omega$, $\nu$), was accomplished following the comprehensive outline provided in the online textbook ``Orbital Mechanics and Astrodynamics''\footnote{\url{https://orbital-mechanics.space/classical-orbital-elements/orbital-elements-and-the-state-vector.html}} by Bryan Weber. The equations and Python code in the online textbook will not be reproduced here. When converting from state vectors to orbital elements, we made the additional calculation of the semi-major axis, $a$, from the specific angular momentum, $h$, and eccentricity, $e$, using
\begin{equation}
a = \frac{h^2}{\mu(1- e^2)},
\end{equation}
where $\mu$ is the standard gravitational parameter, $GM$, for the Sun (1.327 $\times$ 10$^{11}$ km$^3$ s$^{-2}$).

Propagation of the orbit forward in time required determining the true anomaly, $\nu$, 7, 30, and 365 days after the last simulated astrometric measurement. Since the true anomaly does not increase linearly with time, a different angle, the mean anomaly, $M$, was computed first. The first step in this process was to compute the eccentric anomaly, $E$, from the true anomaly at $t$ = 0:
\begin{equation}
E = \mathrm{tan}^{-1}\left(\mathrm{sin}\nu\frac{\sqrt{1 - e^2}}{e + \mathrm{cos}\nu}\right).
\end{equation}
The time of perihelion ($\tau$, in Julian days) was then calculated by equating two definitions of the mean anomaly:
\begin{equation}
M = E - e\mathrm{sin}E
\end{equation}
\begin{equation}
M = n(t - \tau),
\end{equation}
where $n$ is the mean motion and was computed as 2$\pi$ divided by the orbital period, with the orbital period calculated from the standard gravitational parameter and the semi-major axis of the orbit using Newton's version of Kepler's Third Law. The time, $t$, was then incremented by 7, 30, and 365 days, with the mean anomaly computed in each situation.

Working in reverse to calculate the true anomaly required first computing the eccentric anomaly from the mean anomaly. The eccentric and mean anomalies are related via Kepler's equation (A3), but it is transcendental and therefore must be solved numerically. We used the Newton-Raphson method with 10 iterations to determine the eccentric anomaly, and finally calculated the true anomaly using
\begin{equation}
\nu = E + 2\mathrm{arctan}\left(\frac{\beta\mathrm{sin}E}{1-\beta\mathrm{cos}E}\right),
\end{equation}
where $\beta$ = $e/(1+\sqrt{1-e^2})$ \citep{Broucke1973}. The conversion from orbital elements to a state vector was then performed following the recipe described in ``Orbital Mechanics and Astrodynamics.''

\clearpage
\bibliography{ref}{}
\bibliographystyle{aasjournalv7}

\end{document}